\documentclass[aps,onecolumn,showpacs,footinbib]{revtex4}%
\usepackage[latin9]{inputenc}
\usepackage{amssymb}
\usepackage{amsmath}
\usepackage{amscd}
\usepackage{graphicx}
\usepackage{subfigure}
\usepackage{float}
\begin{document}
\draft
\title{State-independent teleportation of an atomic state between two cavities}
\author{Shi-Biao Zheng\thanks{%
E-mail: sbzheng@pub5.fz.fj.cn}}
\address{Department of Electronic Science and Applied Physics\\
Fuzhou University\\
Fuzhou 350002, P. R. China}
\date{\today }

\begin{abstract}
A scheme is presented for the teleportation of an unknown atomic state
between two separated cavities. The scheme involves two
interaction-detection cycles and uses resonantly coupled atoms with an
additional ground state not coupled to the cavity field. Remarkably, the
damping of one basis state is balanced by that of the other basis state and
the state with photon loss in the first interaction-detection cycle is
eliminated by the second cycle. Therefore, the fidelity of teleportation is
independent of the teleported state and insensitive to the atomic
spontaneous emission, cavity decay, and detection inefficiency, which is
obviously in contrast to the original scheme by Bose et al. [Phys. Rev.
Lett. 83 (1999) 5158].
\end{abstract}


\vskip 0.5cm \maketitle

\narrowtext

Quantum teleportation, first proposed by Bennett and colleagues in 1993 [1],
is a means of transporting an unknown quantum state from one place to
another without the requirement to move the particle which carries the
quantum information. As quantum teleportation is one of basic methods of
quantum communication [2] and may be useful in quantum computation [3], it
has attracted considerable attention in recent years. Experimental
realizations of quantum teleportation have been reported using optical
systems with both descrete [4] and continuous variables [5], nuclear
magnetic resonance [6], and trapped ions [7].

The cavity QED system with atoms interacting with quantized electromagnetic
fields is almost an ideal candidate for implementing tasks of quantum
information processing because atoms are suitable for storing information
and photons suitable for transporting information. A number of cavity QED
based schemes have been presented for teleportation of an unknown atomic
state. Most of ealier schemes used atoms as the flying qubits to transfer
quantum information [8]. Therefore, these schemes are not suitable for
long-distance quantum teleportation. The proposal by Bose et al. [9], using
atoms as the stationary qubits and photons as the flying qubits, is
promising for quantum communication. The scheme is based on the detection of
photons leaking out of single-mode cavities in which the atoms are trapped.
Using more complex experimental setups related schemes for quantum
teleportation [10-12] and entanglement engineering have also been presented
[13-17].

In the scheme of Ref. [9], the final state of the second atom is a mixed
state, which is different from the initial state of the first atom, and the
fidelity depends on the teleported state. Furthermore, the fidelity is
affected by cavity decay and imperfect photodetection. This is due to the
fact that the two basis states are not equally damped and the state with two
photons emitted is not eliminated. In this paper, we propose a scheme
involving two interaction-detection cycles using resonantly coupled atoms
with an additional ground state not coupled to the cavity field. In our
scheme, the basis states are equally damped and the state with photon loss
in the first interation-detection cycle is eliminated by the second cycle.
Therefore, the fidelity is state-independent and insensitive to cavity
decay, atomic spontaneous emission, and detection inefficiency. Futhermore,
our scheme uses resonant atom-cavity interaction, instead of Raman coupling
used in Ref. [9], and thus the interaction time is greatly shortened, which
is important for suppressing decoherence and improving the success
probability. Our scheme allows high-fidelity teleportation of an unknown
atomic state without using a more complex experimental setup.

The atoms have one excited state $\left| e\right\rangle $ and two ground
states $\left| g\right\rangle $ and $\left| f\right\rangle $, as shown in
Fig. 1. The quantum information is encoded in the ground states $\left|
f\right\rangle $ and $\left| g\right\rangle $. The transition $\left|
g\right\rangle \rightarrow \left| e\right\rangle $ is resonantly coupled to
the cavity mode. The transition $\left| e\right\rangle \rightarrow \left|
f\right\rangle $ is dipole-forbidden. The setup is shown in Fig. 2. Two
distant atoms are trapped in two separate single-mode optical cavities,
respectively. Photons leaking out of the cavities are mixed on a
beam-splitter, which destroys which-path information. Then the photons are
detected by two photodetectors. We here assume that the cavities are one
sided so that the only photon leakage occurs through the sides of the
cavities facing the beam-splitter.

The whole procedure of our scheme is composed of quantum information
splitting and recombination. Each atom is first entangled with the
corresponding cavity mode via resonant interaction. The detection of one
photon leaking out of the cavities and passing through the beam-splitter
corresponds to measurement of the joint state of the two cavities, which
collapses the two distant atoms to an entangled state. After the first
interaction-detection cycle, the quantum information initially encoded in
atom 1 is shared by the two atoms. The process is referred to as
entanglement swapping [18]. During the second interaction, if the atoms emit
one photon the quantum state of atom 1 is transferred to the cavity fields.
The detection of the photon enables the splitted quantum information to be
recombined and completely encoded in atom 2 [19].

Assume that the atom (atom 1), whose state is to be teleported, is initially
in the state
\begin{equation}
\left| \phi _1\right\rangle =c_f\left| f_1\right\rangle +c_g\left|
g_1\right\rangle ,
\end{equation}
where $c_f$ and $c_g$ are unknown coefficients. Both the two cavities are
initially in the vacuum state $\left| 0\right\rangle $. The first step is
the transfer of one photon to the cavity through a half-cycle of the vacuum
Rabi-oscillation of the atom-cavity system. The vacuum Rabi half-cycle is
initiated by exciting the state $\left| g_1\right\rangle $ to $\left|
e_1\right\rangle $. This leads to
\begin{equation}
\left| \phi _1^{^{\prime }}\right\rangle =c_f\left| f_1\right\rangle
+c_g\left| e_1\right\rangle .
\end{equation}
The emission or non-emission of a photon depends on whether the initial
state is $\left| g_1\right\rangle $ or $\left| f_1\right\rangle $, providing
the essential tool for generating entanglement between the atom and cavity
field. This is distinguished with the scheme of Ref. [9], in which two
ground atomic states are coupled to the cavity field via Raman process and
atom 1 is disentangled with the cavity field after the atom-cavity
interaction. The atom (atom 2), to receive the teleported state, is
initially prepared in the state
\begin{equation}
\left| \phi _2\right\rangle =\frac 1{\sqrt{2}}(\left| e_2\right\rangle
+\left| f_2\right\rangle ).
\end{equation}
In Ref. [9], atom 2 is initially in a ground state and the entanglement
between atom 2 and cavity 2 is obtained after a Rabi quater-cycle, which is
obviously in contrast with the present case. The aim of using the initial
state $\left| \phi _2\right\rangle $ is to let the basis states $\left|
g_1\right\rangle \left| f_2\right\rangle $ and $\left| f_1\right\rangle
\left| g_2\right\rangle $ be equally damped, as shown below. This allows the
effect of decoherence in cavity 1 is balanced by that in cavity 2.

In the interaction picture, the Hamiltonian in each cavity is
\begin{equation}
H_j=g(a_jS_j^{+}+a_j^{+}S_j^{-}),
\end{equation}
where $S_j^{+}=\left| e_j\right\rangle \left\langle g_j\right| $ and $%
S_j^{-}=\left| g_j\right\rangle \left\langle e_j\right| $ are the raising
and lowering operators of the jth (j=1,2) atom, $a_j$ and $a_j^{+}$ are the
annihilation and creation operators of the jth cavity mode, and g is the
atom-cavity coupling strength. The Hamiltonian of Eq. (4) does not include
the effects of the atomic spontaneous emission and cavity decay. Under the
condition that no photon is detected either by the atomic spontaneous
emission or by the leakage through the cavity mirror, the evolution of the
system is governed by the conditional Hamiltonian
\begin{equation}
H_{con,j}=H_j-i\frac \kappa 2a_j^{+}a_j-i\frac \Gamma 2\left|
e_j\right\rangle \left\langle e_j\right| ,
\end{equation}
where $\kappa $ is the cavity decay rate and $\Gamma $ is the atomic
spontaneous emission rate. The time evolution for the state $\left|
e_j\right\rangle \left| 0_j\right\rangle $ is
\begin{equation}
\left| e_j\right\rangle \left| 0_j\right\rangle \rightarrow e^{-(\kappa
+\Gamma )t/4}\{[\cos (\beta t)+\frac{\kappa -\Gamma }{4\beta }\sin (\beta
t)]\left| e_j\right\rangle \left| 0_j\right\rangle -i\frac g\beta \sin
(\beta t)\left| g_j\right\rangle \left| 1_j\right\rangle \},
\end{equation}
where
\begin{equation}
\beta =\sqrt{g^2-(\kappa -\Gamma )^2/16}.
\end{equation}
After an interaction time $t_1$ given by $\tan (\beta t_1)=4\beta /(\Gamma
-\kappa )$, the whole system evolves to
\begin{eqnarray}
\left| \psi _1\right\rangle &=&\frac 1{\sqrt{2}}\{c_f\left| f_1\right\rangle
\left| 0_1\right\rangle -ic_g\frac g\beta e^{-(\kappa +\Gamma )t_1/4}\sin
(\beta t_1)\left| g_1\right\rangle \left| 1_1\right\rangle \} \\
&&\ \{\left| f_2\right\rangle \left| 0_2\right\rangle -i\frac g\beta
e^{-(\kappa +\Gamma )t_1/4}\sin (\beta t_1)\left| g_2\right\rangle \left|
1_2\right\rangle \}.  \nonumber
\end{eqnarray}
Unlike the scheme of Ref. [9], the state of atom 1 is not transferred to
cavity 1, and atom 2 and cavity 2 are not prepared in a maximally entangled
state.

Now we perform the transformation:

\begin{equation}
\left| f_j\right\rangle \rightarrow \left| g_j\right\rangle ;\left|
g_j\right\rangle \rightarrow -\left| f_j\right\rangle .
\end{equation}
This leads to
\begin{eqnarray}
\left| \psi _2\right\rangle &=&\frac 1{\sqrt{2}}\{c_f\left| g_1\right\rangle
\left| 0_1\right\rangle +ic_g\frac g\beta e^{-(\kappa +\Gamma )t_1/4}\sin
(\beta t_1)\left| f_1\right\rangle \left| 1_1\right\rangle \} \\
&&\ \{\left| g_2\right\rangle \left| 0_2\right\rangle +i\frac g\beta
e^{-(\kappa +\Gamma )t_1/4}\sin (\beta t_1)\left| f_2\right\rangle \left|
1_2\right\rangle \}.  \nonumber
\end{eqnarray}
After the transformation the atom-cavity interaction is frozen since $%
H_j\left| \psi _2\right\rangle =0$. Now we waits for the photodetectors to
click. The registering of a click at one of the photodetectors corresponds
to the action of the jump operators $(a_1\pm a_2)/\sqrt{2}$ on the state $%
\left| \psi _2\right\rangle $, where ''+'' corresponds to the detection of
the photon at the photodetector D$_{+}$, while ''-'' corresponds to the
detection of the photon at D$_{-}$. The system is then projected to
\begin{eqnarray}
\left| \psi _3\right\rangle &=&ic_g\frac g{2\beta }e^{-(\kappa +\Gamma
)t_1/4-\kappa \tau _1/2}\sin (\beta t_1)\left| f_1\right\rangle \left|
0_1\right\rangle \left| g_2\right\rangle \left| 0_2\right\rangle \\
&&\pm ic_f\frac g{2\beta }e^{-(\kappa +\Gamma )t_1/4-\kappa \tau _1/2}\sin
(\beta t_1)\left| g_1\right\rangle \left| 0_1\right\rangle \left|
f_2\right\rangle \left| 0_2\right\rangle  \nonumber \\
&&-c_g\frac{g^2}{2\beta ^2}e^{-(\kappa +\Gamma )t_1/2-\kappa \tau _1}\sin
^2(\beta t_1)\left| f_1\right\rangle \left| f_2\right\rangle (\left|
0_1\right\rangle \left| 1_2\right\rangle \pm \left| 1_1\right\rangle \left|
0_2\right\rangle ),  \nonumber
\end{eqnarray}
where $\tau _1$ is the waiting time. In comparison with the scheme of Ref.
[9], after the detection of the photon atom 1 is entangled with atom 2 and
the cavity modes, and the two basis states $\left| f_1\right\rangle \left|
g_2\right\rangle $ and $\left| g_1\right\rangle \left| f_2\right\rangle $
are equally damped. Then we wait for another time $\tau _2$. Suppose that no
photon is detected during this period. Due to the cavity decay the system
evolves to
\begin{eqnarray}
\left| \psi _4\right\rangle &=&ic_g\frac g{2\beta }e^{-(\kappa +\Gamma
)t_1/4-\kappa \tau _1/2}\sin (\beta t_1)\left| f_1\right\rangle \left|
0_1\right\rangle \left| g_2\right\rangle \left| 0_2\right\rangle \\
&&\ \ \pm ic_f\frac g{2\beta }e^{-(\kappa +\Gamma )t_1/4-\kappa \tau
_1/2}\sin (\beta t_1)c_f\left| g_1\right\rangle \left| 0_1\right\rangle
\left| f_2\right\rangle \left| 0_2\right\rangle  \nonumber \\
&&\ \ -c_g\frac{g^2}{2\beta ^2}e^{-(\kappa +\Gamma )t_1/2-\kappa (\tau
_1+\tau _2/2)}\sin ^2(\beta t_1)\left| f_1\right\rangle \left|
f_2\right\rangle (\left| 0_1\right\rangle \left| 1_2\right\rangle \pm \left|
1_1\right\rangle \left| 0_2\right\rangle ).  \nonumber
\end{eqnarray}
If $\tau _2$ is long enough so that $e^{-\kappa \tau _2/2}\ll 1$ the last
term of $\left| \psi _4\right\rangle $ can be discarded. This leads to
\begin{equation}
\left| \psi _4\right\rangle =i\frac g{2\beta }e^{-(\kappa +\Gamma
)t_1/4-\kappa \tau _1/2}\sin (\beta t_1)(c_g\left| f_1\right\rangle \left|
0_1\right\rangle \left| g_2\right\rangle \left| 0_2\right\rangle \pm
c_f\left| g_1\right\rangle \left| 0_1\right\rangle \left| f_2\right\rangle
\left| 0_2\right\rangle ).
\end{equation}

Then we sequentially perform the following transformations on atom 1: $%
\left| g_1\right\rangle \rightarrow \left| e_1\right\rangle $ and $\left|
f_1\right\rangle \rightarrow \left| g_1\right\rangle $. Meanwhile we excite
the state $\left| g_2\right\rangle $ to $\left| e_2\right\rangle $. This
leads to
\begin{equation}
\left| \psi _5\right\rangle =i\frac g{2\beta }e^{-(\kappa +\Gamma
)t_1/4-\kappa \tau _1/2}\sin (\beta t_1)(c_g\left| g_1\right\rangle \left|
0_1\right\rangle \left| e_2\right\rangle \left| 0_2\right\rangle \pm
c_f\left| e_1\right\rangle \left| 0_1\right\rangle \left| f_2\right\rangle
\left| 0_2\right\rangle ).
\end{equation}
Due to the transformations each atom interacts with the corresponding cavity
mode again. Suppose that no photon is detected during the interaction the
evolution of the system is
\begin{eqnarray}
\left| \psi _6\right\rangle &=&ic_g\frac g{2\beta }e^{-(\kappa +\Gamma
)(t_1+t_2)/4-\kappa \tau _1/2}\sin (\beta t_1)\left| g_1\right\rangle \left|
0_1\right\rangle \{[\cos (\beta t_2) \\
&&\ +\frac{\kappa -\Gamma }{4\beta }\sin (\beta t_2)]\left| e_2\right\rangle
\left| 0_2\right\rangle -i\frac g\beta \sin (\beta t_2)\left|
g_2\right\rangle \left| 1_2\right\rangle \}  \nonumber \\
&&\pm ic_f\frac g{2\beta }e^{-(\kappa +\Gamma )(t_1+t_2)/4-\kappa \tau
_1/2}\sin (\beta t_1)\{[\cos (\beta t_2)  \nonumber \\
&&+\frac{\kappa -\Gamma }{4\beta }\sin (\beta t_2)]\left| e_1\right\rangle
\left| 0_1\right\rangle -i\frac g\beta \sin (\beta t_2)\left|
g_1\right\rangle \left| 1_1\right\rangle \}\left| f_2\right\rangle \left|
0_2\right\rangle ,  \nonumber
\end{eqnarray}
where $t_2$ is the interaction time. The registering of a click at one of
the photodetectors at some moment projects atom 2 to
\begin{equation}
\left| \varphi \right\rangle =c_g\left| g_2\right\rangle \pm c_f\left|
f_2\right\rangle ,
\end{equation}
with atom 1 left in the state $\left| g_1\right\rangle $ and the two cavity
modes left in the vacuum state $\left| 0_1\right\rangle \left|
0_2\right\rangle $. Here ''+'' corresponds to the detection of two photons
at the same photodetector during the two interaction-detection cycles, while
''-'' corresponds to the detection of the photons at different
photodetectors. If the two photons are detected in the same photodetector
atom 2 is just in the initial state of atom 1. If the two photons are
detected at different photodetectors we perform the rotation $\left|
f_2\right\rangle \rightarrow -\left| f_2\right\rangle $ to reconstruct the
initial state of atom 1.

If one wishes to wait a time $t_d$ during the second stage, the probability
of success is

\begin{eqnarray}
P &=&\frac 12e^{-(\kappa +\Gamma )t_1/2}\{1-e^{-(\kappa +\Gamma )t_d/2}[\cos
^2(\beta t_d) \\
&&+\frac{(\kappa -\Gamma )^2+4g^2}{4\beta ^2}\sin ^2(\beta t_d)+\frac{\kappa
-\Gamma }{4\beta }\sin (2\beta t_d)]\}.  \nonumber
\end{eqnarray}
If $t_d$ is long enough so that $e^{-(\kappa +\Gamma )t_d/2}\ll 1$ the
success probability is $P=e^{-(\kappa +\Gamma )t_1/2}/2$. The success
probability increases as the needed interaction time $t_1$ decreases.

Due to the imperfection of the photodetectors, there is a probability that
two photons have leaked out of the cavities but only one photon is detected
during the first interaction-detection cycle, which leads to the state $%
\left| f_1\right\rangle \left| f_2\right\rangle \left| 0_1\right\rangle
\left| 0_2\right\rangle $. In this case no photon is emitted during the
second cycle and the event is discarded. The scheme is conditional upon the
detection of two emitted photons. If one of the emissions is not detected,
the scheme fails and the procedure restarts. Therefore, the imperfection of
the photodetectors does not affect the fidelity of the teleported state. Set
the detection efficiency to be $\eta $. Then the success probability is $%
P^{^{\prime }}=\eta ^2P$.

Because of imperfect timing of the interaction time $t_1$, atom 2 is finally
in the mixed state

\begin{equation}
\rho =(\left| \varepsilon _1\right| ^2+2\left| \varepsilon _{2,\pm }\right|
^2+2\left| \varepsilon _{3,\pm }\right| ^2)^{-1}[(\left| \varepsilon
_1\right| ^2+\left| \varepsilon _{2,\pm }\right| ^2)\left| \varphi
^{^{\prime }}\right\rangle \left\langle \varphi ^{^{\prime }}\right|
+(\left| \varepsilon _{2,\pm }\right| ^2+2\left| \varepsilon _{3,\pm
}\right| ^2)\left| g_2\right\rangle \left\langle g_2\right| ],
\end{equation}
where
\begin{equation}
\left| \varphi ^{^{\prime }}\right\rangle =(\left| \varepsilon _1\right|
^2+\left| \varepsilon _{2,\pm }\right| ^2)^{-1/2}[\varepsilon _1(\pm
c_g\left| g_2\right\rangle \pm c_f\left| f_2\right\rangle )+\varepsilon
_{2,\pm }\left| e_2\right\rangle ],
\end{equation}
and
\begin{eqnarray}
\varepsilon _1 &=&BEe^{-\kappa \tau _1/2}, \\
\varepsilon _{2,\pm } &=&(c_g\pm c_f)ACDE,  \nonumber \\
\varepsilon _{3,\pm } &=&(c_g\pm c_f)ACE^2,  \nonumber \\
A &=&e^{-(\kappa +\Gamma )(t_1+\delta t_1)/4}\{[\cos [\beta (t_1+\delta
t_1)]+\frac{\kappa -\Gamma }{4\beta }\sin [\beta (t_1+\delta t_1)]\},
\nonumber \\
B &=&i\frac g\beta e^{-(\kappa +\Gamma )(t_1+\delta t_1)/4}\sin [\beta
(t_1+\delta t_1)],  \nonumber \\
C &=&-i\frac g\beta e^{-(\kappa +\Gamma )\tau _1/4}\sin (\beta \tau _1),
\nonumber \\
D &=&e^{-(\kappa +\Gamma )t_2/4}[\cos (\beta t_2)+\frac{\kappa -\Gamma }{%
4\beta }\sin (\beta t_2)],  \nonumber \\
E &=&-i\frac g\beta e^{-(\kappa +\Gamma )t_2/4}\sin (\beta t_2).  \nonumber
\end{eqnarray}
Here $\delta t_1$ is the deviation from the desired interaction time. With
imperfect timing of the interaction time $t_1$ being considered, the
fidelity is
\begin{equation}
F=\frac{\left| \varepsilon _1\right| ^2+(\left| \varepsilon _{2,\pm }\right|
^2+2\left| \varepsilon _{3,\pm }\right| ^2)\left| c_g\right| ^2}{\left|
\varepsilon _1\right| ^2+2\left| \varepsilon _{2,\pm }\right| ^2+2\left|
\varepsilon _{3,\pm }\right| ^2}.
\end{equation}
When the timing is not perfect the fidelity depends on the state to be
teleported.

In order to perform the transformation in Eq. (11) we use a pair of
off-resonant classical fields with the same Rabi frequency $\Omega $ to
drive the transitions $\left| g_j\right\rangle \rightarrow \left|
h_j\right\rangle $ and $\left| f_j\right\rangle \rightarrow \left|
h_j\right\rangle $ , where $\left| h_j\right\rangle $ is an auxiliary
excited state. The two classical fields are detuned from the respective
transitions by the same amount $\delta $. In the case that the detuning $%
\delta $ is much larger than the Rabi frequency $\Omega $ the upper level $%
\left| h_j\right\rangle $ can be adiabatically eliminated and the two
classical fields just induce the Raman transition between the states $\left|
g_j\right\rangle $ and $\left| f_j\right\rangle $ [20]. The Raman coupling
strength is $\lambda =\Omega ^2/\delta $. The time needed to perform the
required transformation is $\pi /2\lambda $. Under the condition $\lambda
\gg g$, the atom-cavity interaction can be neglected during this
transformation. Set $\Omega =3\times 10^2g$ and $\delta =10\Omega $. During
this transformation the probability that each atom exchanges an excitation
with the cavity mode is on the order of $(g\pi /2\lambda )^2\simeq
2.74\times 10^{-3}$.

The required atomic level configuration can be achieved in Cs. The hyperfine
levels $\left| F=4,m=-1\right\rangle $ and $\left| F=4,m=0\right\rangle $ of
$6S_{1/2}$ can act as the ground states $\left| g\right\rangle $ and $\left|
f\right\rangle $, respectively, while the hyperfine levels $\left|
F^{^{\prime }}=5,m^{^{\prime }}=0\right\rangle $ and $\left| F^{^{\prime
}}=5,m^{^{\prime }}=-1\right\rangle $ of $5^2P_{3/2}$ can act as the excited
states $\left| e\right\rangle $ and $\left| h\right\rangle $, respectively.
In a recent cavity QED experiment with Cs atoms trapped in an optical
cavity, the corresponding atom-cavity coupling strength is $g=2\pi \times
34MHz$ [21]. The decay rates for the atomic excited states and the cavity
mode are $\Gamma =2\pi \times 2.6MHz$ and $\kappa =2\pi \times 4.1MHz$,
respectively. The required interaction time $t_1$ is about $7.4\times
10^{-3}\mu s$. The waiting times $\tau _1$ and $\tau _2$ are on the order of
$2/\kappa \simeq 7.8\times 10^{-2}\mu s$ and $20/\kappa \simeq 7.8\times
10^{-1}\mu s$, respectively. The waiting time $t_d$ during the second stage
is on the order of $20/(\kappa +\Gamma )\simeq 4.8\times 10^{-1}\mu s$. The
total time needed to complete the teleportation is on the order of $1.35\mu
s $. Set $c_f=c_g=\frac 1{\sqrt{2}}$, $\delta t_1=0.05t_1$, and $\eta =0.6$
[9]. Then the success probability and fidelity are about 0.15 and 0.998,
respectively. The present scheme works in the Lamb-Dicke regime, i.e., the
spatial extension of the atomic wave function should be much smaller than
the wavelength of the light fields. In a recent experiment [22], the
localization to the Lamb-dicke limit of the axial motion was demonstrated
for a single atom trapped in an optical cavity.

In summary, we have proposed a scheme for long-distance teleportation of the
state of an atom trapped in an optical cavity to a second atom trapped in
another distant optical cavity. The scheme involves two
interaction-detection cycles and uses resonant atoms with an additional
ground state not coupled to the cavity field. The distinct advantage of our
scheme is that the teleportation fidelity is state-independent and
insensitive to decoherence and imperfect photodetection in principle.

This work was supported by the National Natural Science Foundation of China
under Grant No. 10674025 and the Doctoral Foundation of the Ministry of
Education of China under Grant No. 20070386002.

\begin{figure}[C]
\includegraphics[width=0.2\columnwidth]{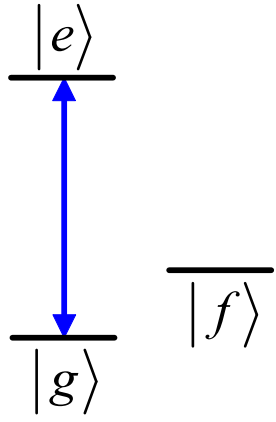}
\caption{The level configuration of the atoms. The transition
$\left| g\right\rangle \rightarrow \left| e\right\rangle $ is
resonantly coupled to the cavity mode and the additional ground
state $\left| f\right\rangle $ is not coupled to the cavity mode.}
\end{figure}

\begin{figure}[C]
\includegraphics[width=0.5\columnwidth]{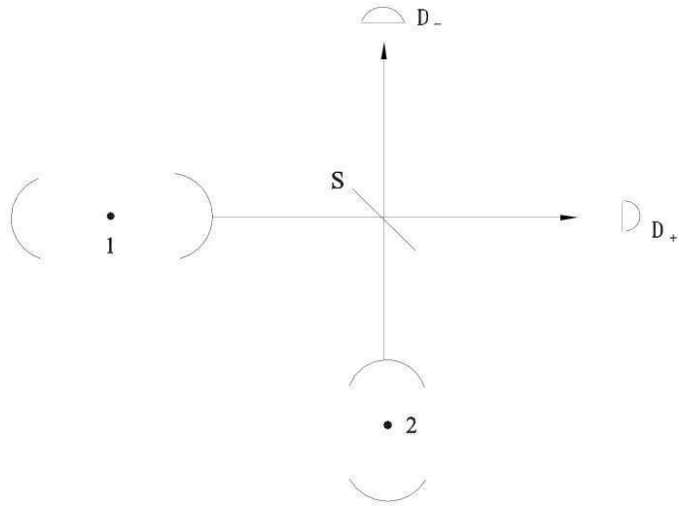}
\caption{The experimental setup. Two distant atoms are trapped in
separate cavities. Photons leak through the sides of the cavities
facing the
beam-splitter S and then are detected by the photodetectors D$_{+}$ and D$%
_{-}$.}
\end{figure}
\end{document}